\documentclass[aps,pra,twocolumn]{revtex4-1}
\usepackage{graphicx}
\usepackage{amsmath}
\usepackage{amsfonts}
\usepackage{amssymb}
\usepackage{epstopdf}
\usepackage{color}
\usepackage{bm}% bold math
\newcommand{\ket}[1]{|#1\rangle}
\newcommand{\bra}[1]{\langle #1|}

\begin{document}
\today
\title{Electric nonadiabatic geometric entangling gates on spin qubits}
\author{Vahid Azimi Mousolou\footnote{Electronic address: v.azimi@sci.ui.ac.ir}}
\affiliation{Department of Mathematics, Faculty of Science, University of Isfahan, Box 81745-163 Isfahan, Iran}
\begin{abstract}
Producing and maintaining entanglement reside at the heart of the optimal construction of quantum operations and are fundamental issues in the realization of universal quantum computation. We here introduce a setup of spin qubits that allows for geometric implementation of entangling gates between the register qubits with any arbitrary entangling power. We show this by demonstrating a circuit through a spin chain, which performs universal nonadiabatic holonomic two-qubit entanglers. The proposed gates are all electric and geometric, which would help to realize fast and robust entangling gates on spin qubits. This family of entangling gates contains gates that are as efficient as the CNOT gate in quantum algorithms.
We examine the robustness of the circuit to some extent.
\end{abstract}
\maketitle
\section{Introduction}

Entanglement is the main ingredient of various insights into quantum information processing such as 
quantum cryptography, superdense coding, quantum teleportation, quantum error correction,
and efficient quantum computation \cite{steane98, DiVincenzo2000, Horodecki2009, niellsen10}. In a sense entanglement is a necessary asset to realize quantum computing. 
This primary quantum physical phenomenon is produced by non-local unitary quantum evolutions, i.e., quantum gates acting 
on the state space of the multi-qubit system that cannot be decomposed into a product of single-qubit gates \cite{niellsen10}. These non-local quantum gates are typically referred to as entanglers. Among entanglers, two-qubit entanglers play a central role in the optimal construction of universal quantum operations. Therefore, there has been much efforts devoted to the physical realization of two-qubit entangling gates in an efficient way, i.e., a way that is fast and robust.  

Inherent robustness of non-Abelian quantum geometric phases \cite{Pachos2001, sjoqvist2008, Solinas2012} distinguishes a prominent way for implementation of quantum gates.  
This approach, which is known as holonomic quantum computation, was originally conceived \cite{zanardi99} based on the adiabatic non-Abelian geometric phase \cite{wilczek84}. Recently, this approach has been generalized \cite{sjoqvist12} based on the nonadiabatic non-Abelian geometric phase \cite{anandan88}. 
The generalized approach allows us to combine the necessary components for realization of quantum processors, i.e., robustness, universality and speed \cite{sjoqvist12, Johansson12, sjoqvist16}. It has been shown that the idea of nonadiabatic holonomic quantum computation can be incorporated with decoherence free subspaces \cite{Xu2012, Xu2014a, Liang2014, Xue2015, Zhou2015, Xue2016}, noiseless subsystems \cite{Zhang2014}, and dynamical decoupling \cite{Xu2014b}.
Non-adiabatic holonomic quantum computation has been realized in different experimental settings, such as NMR \cite{Feng2013}, superconducting transmon \cite{Abdumalikov2013}, and nitrogen-vacancy centers in diamond \cite{Arroyo-Camejo2014, Zu2014}. Nonetheless, the need for control of complicated interactions between qubit systems has made it a challenging task to realize geometric two-qubit entangling gates. In some of these experimental settings only nonadiabatic holonomic single-qubit gates have been realized.

In this paper we introduce a three-body XY spin-chain system to realize nonadiabatic holonomic two-qubit entangling gates between two spin qubit registers.
We show that this system permits for practical implementation of a fast and geometric family of two-qubit entanglers with arbitrary entangling power.  The gates are accomplished by electrical control of inter-qubit exchange couplings.  We examine the geometric and entangling nature of the gates. We show that the proposed family of nonlocal gates contains all types of entanglers including perfect entangler, special perfect entangler, and entanglers that are as efficient as the CNOT gate in quantum algorithms.  We show that only the anisotropic XY interaction between qubits is sufficient to realize all these different types of entanglers. The proposed system can be realized with three coplanar quantum dot spin qubits within an in-plane electric field. However, the system Hamiltonian is a general one that can be achieved with different physical systems actively considered for realization of quantum processors.

We begin with introducing our spin model system and the corresponding dynamics in Sec.~\ref{Model system}. In Sec.~\ref{Two-qubit entangling gate}, we establish a circuit to realize nonlocal geometric two-qubit gates on register spin qubits, and examine the entanglement characteristics of these gates. We continue in Sec.~\ref{geometric interpretation}, to shed light on the geometric nature and feasibility of the proposed gates. The robustness of the gates are studied in Sec.~\ref{Robustness}. The paper is summarized in Sec.~\ref{summary}.

\section{Model system}
\label{Model system}

\begin{figure}[h]
\centering
\includegraphics[width=65mm,height=16mm]{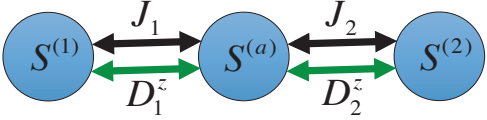}
\caption{(Color online) Two register spin qubits labeled by $S^{(1)}$ and $S^{(2)}$ are coupled through an intermediate ancilla spin qubit labeled as $S^{(a)}$. The ancilla qubit allows us to evolve two register qubits by means of nonadiabatic quantum holonomy and generate geometric entanglement between the two register spin qubits.}
\label{fig:3-spin-chain}
\end{figure}

The system that we consider here is a three-body spin-chain system described effectively by the Hamiltonian
\begin{eqnarray}
H_{\text{eff}}=H_{\text{XY}}+H_{\text{DM}}
\label{eq:eff-H}
\end{eqnarray}
where
\begin{eqnarray}
H_{\text{XY}}= J_{1}[S_{x}^{(1)}S_{x}^{(a)}+S_{y}^{(1)}S_{y}^{(a)}]+J_{2}[S_{x}^{(a)}S_{x}^{(2)}+S_{y}^{(a)}S_{y}^{(2)}]\nonumber\\
\end{eqnarray}
is the anisotropic XY interaction Hamiltonian with exchange coupling strength $J_{k}$, $k=1,2$, and
\begin{eqnarray}
H_{\text{DM}}= D^{z}_{1}[S_{x}^{(1)}S_{y}^{(a)}-S_{y}^{(1)}S_{x}^{(a)}]+D^{z}_{2}[S_{x}^{(a)}S_{y}^{(2)}-S_{y}^{(a)}S_{x}^{(2)}]\nonumber\\
\end{eqnarray}
is the antisymmetric Dzyalozhinsky-Moriya spin-orbit interaction term with  exchange coupling strength $D_{k}^{z}$. This system can be realized for instance with three coplanar quantum dot spin qubits in the $xy$ plane within an in-plane electric field \cite{imamoglu1999, trif10}.  As illustrated in Fig. \ref{fig:3-spin-chain}, in this system we assume that two register spin qubits are coupled through an intermediate ancilla spin qubit. 

We assume that exchange parameters are turned on and off by a common time-dependent scaling function $\Omega(t)$, i.e., $J_{k}=\Omega(t)\mathcal{J}_{k}$ and $D_{k}^{z}=\Omega(t)\mathcal{D}_{k}^{z}$, $k=1,2$. Note that since the exchange parameters $J_{k}$ and $D_{k}^{z}$ are proportional to the inter-dot hopping terms \cite{trif10}, they can be efficiently controlled with time-dependent electric gate voltages. Using a particular index order ($a$, 1, 2), we can put the time-dependent effective Hamiltonian in the following block off-diagonal form
\begin{eqnarray}
H_{\text{eff}}(t)=\Omega(t)[S_{+}^{(a)}\otimes W + h.c.],
\label{t-Hamiltonian}
\end{eqnarray}
where
\begin{eqnarray}
W=\left(
\begin{array}{cccc}
 0 &   0 & 0 & 0  \\
\alpha_{2}  &  0 & 0 & 0  \\
 \bar{\alpha}_{1} & 0 & 0 & 0 \\
  0 &  \bar{\alpha}_{1} & \alpha_{2}  & 0
\end{array}
\right)
\end{eqnarray}
with $\alpha_{1}=\frac{\hbar}{2}(\mathcal{J}_{1}+i\mathcal{D}_{1}^{z})$ and $\alpha_{2}=\frac{\hbar}{2}(\mathcal{J}_{2}+i\mathcal{D}_{2}^{z})$ is the  time-independent exchange matrix operator in the computational basis acting on the register two-qubit space. Here $S_{+}^{(a)}=S_{x}^{(a)}+iS_{y}^{(a)}=\hbar\ket{0}\bra{1}$ is the raising operator for the ancilla spin qubit with $\ket{0}$ and $\ket{1}$ representing spin-up and spin-down states.

Doing a singular value decomposition of  $W$,  we obtain $W=V_{0}TV^{\dagger}_{1}$ with
\begin{eqnarray}
T=\left(
\begin{array}{cccc}
0 &   0 & 0 & 0  \\
0  &  0 & 0 & 0  \\
0 & 0 & \omega & 0 \\
0 & 0 & 0  & \omega
\end{array}
\right)
\end{eqnarray}
and 
\begin{eqnarray}
V_{0}&=&\left(
\begin{array}{cccc}
1 &   0 & 0 & 0  \\
0  &  e^{i\phi_{1}}\cos\theta & 0  & e^{i\phi_{2}}\sin\theta  \\
0 &- e^{-i\phi_{2}}\sin\theta & 0 & e^{-i\phi_{1}}\cos\theta\\
0 & 0 & 1  & 0
\end{array}
\right)\nonumber\\
V_{1}&=&\left(
\begin{array}{cccc}
0 &   0 & 0 &1  \\
0  &  e^{i\phi_{2}}\sin\theta & e^{i\phi_{1}}\cos\theta  & 0  \\
0 &- e^{-i\phi_{1}}\cos\theta  &e^{-i\phi_{2}}\sin\theta& 0 \\
1 & 0 & 0  & 0
\end{array}
\right),
\end{eqnarray}
where $\omega=\sqrt{|\alpha_{1}|^{2}+|\alpha_{2}|^{2}}$ and 
\begin{eqnarray}
\frac{\alpha_{1}}{\omega}= e^{i\phi_{1}}\cos\theta \ \ \ \& \ \ \ \frac{\alpha_{2}}{\omega}= e^{i\phi_{2}}\sin\theta.
\label{exchange-coupling-constants}
\end{eqnarray}
Using this decomposition, we obtain the time evolution operator as  
\begin{eqnarray}
\mathcal{U}(t, 0)&=&\exp[-\frac{i}{\hbar}\int_{0}^{t}H_{\text{eff}}(s)ds]\nonumber\\
&=&\cos[a_{t}\left(\begin{array}{cccc}
0 &   W \\
W^{\dagger}  &  0
\end{array}\right)]-i\sin[a_{t}\left(\begin{array}{cccc}
0 &   W \\
W^{\dagger}  &  0
\end{array}\right)]\nonumber\\
&=&\sum_{k=0}^{1}\sum_{l=0}^{1}i^{|k-l|}\ket{l}\bra{k}\otimes V_{l}\cos(a_{t}T+\frac{|k-l|\pi}{2}\hat{1}) V_{k}^{\dagger},\nonumber\\
\label{t-evolution}
\end{eqnarray}
where $a_{t}=\int_{0}^{t}\Omega(s)ds$ and $\hat{1}$ is the $4\times 4$ identity matrix. Note that in Eq. (\ref{t-evolution}), the left and right hand sides of each tensor product, respectively, act on the ancilla qubit and the register two-qubit system.

\section{Two-qubit entangling gate}
\label{Two-qubit entangling gate}
In what follows, we show that the above system allows for entangling gates between register qubits, provided the ancilla qubit is initialized in the $\ket{0}$ state. As the ancilla qubit is initialized in the state $\ket{0}$, the three qubit  system would be initially in the subspace 
\begin{eqnarray}
\mathcal{H}_{0}=\text{Span}\{\ket{000}, \ket{001}, \ket{010}, \ket{011}\}=\ket{0}\otimes\mathcal{H}_{r}
\end{eqnarray}
of the eight-dimensional three qubit Hilbert space $\mathcal{H}$. Here, $\mathcal{H}_{r}$ denotes the computational space of the register two-qubit system, while $\ket{0}$ at the first site represents the state of the ancilla qubit.

\begin{figure}[h]
\centering
\includegraphics[width=80mm,height=15mm]{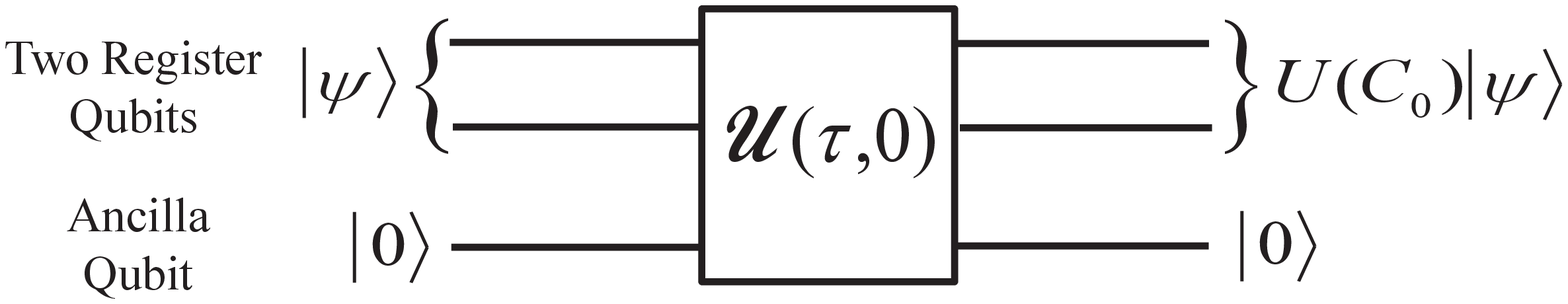}
\caption{Schematic circuit diagram of the geometric two-qubit entangling gate $U(C_{0})$.}
\label{fig:entangler}
\end{figure}

Let us now consider the Schr\"odinger time evolution of the subspace $\mathcal{H}_{0}(t)$ started at $\mathcal{H}_{0}(0)=\mathcal{H}_{0}$, i.e., the path
\begin{eqnarray}
C_{0}: [0, \tau]\ni t \rightarrow \mathcal{H}_{0}(t),
\end{eqnarray}
where each state $\ket{\psi(t)}\in\mathcal{H}_{0}(t)$ is a solution of the time-dependent Schr\"odinger equation at time $t$ for a given initial state in $\mathcal{H}_{0}$. Suppose $\mathcal{H}_{0}(t)$ evolves in a cyclic manner, i.e.,  there is a time $\tau$ such that 
 $\mathcal{H}_{0}(\tau)= \mathcal{H}_{0}(0)=\mathcal{H}_{0}$. The expression for the time evolution operator given by  Eq. (\ref{t-evolution}), implies that the path $C_{0}$ would be cyclic if we choose coupling constants and envelop function $\Omega(t)$ in the time-dependent effective Hamiltonian in Eq. (\ref{t-Hamiltonian}), as well as the final time $\tau$ in a way that  $a_{\tau}\omega$ is an integer multiple of $\pi$. In particular if we consider $a_{\tau}\omega=(2n+1)\pi$, then each given initial state $\ket{0}\otimes\ket{\psi}$ in $\mathcal{H}_{0}$, where $\ket{\psi}\in\mathcal{H}_{r}$ is an initial state for the two register qubits, evolves through the loop $C_{0}$ into the final state 
\begin{eqnarray}
\mathcal{U}(\tau, 0)[\ket{0}\otimes\ket{\psi}]=\ket{0}\otimes U(C_{0})\ket{\psi}\in \mathcal{H}_{0}.
\end{eqnarray}
In the computational basis the unitary $U(C_{0})$ is the following two-qubit gate 
\begin{eqnarray}
U(C_{0})&=&V_{0}\left(
\begin{array}{cccc}
1  &   0 & 0 & 0  \\
0  &  1 & 0  & 0  \\
0  & 0 & -1 & 0\\
0 & 0 & 0 & -1
\end{array}
\right)V_{0}^{\dagger}\nonumber\\
&=&\left(
\begin{array}{cccc}
1 &   0 & 0 & 0  \\
0  &  \cos2\theta & -e^{i(\phi_{1}+\phi_{2})}\sin2\theta & 0  \\
0 & -e^{-i(\phi_{1}+\phi_{2})}\sin2\theta & -\cos2\theta & 0\\
0 & 0 & 0  & -1
\end{array}
\right),\nonumber\\
 \label{eq:entangler}
\end{eqnarray}
which manipulates only register two-qubit states. This procedure is shown in Fig. \ref{fig:entangler}.

We notice that the unitary operator $U(C_{0})$ in Eq. ( \ref{eq:entangler}) introduces an entangling gate between two register spin qubits. In order to clarify this point, below we examine some entangling characteristic of  $U(C_{0})$.

\begin{figure}[h]
\centering
\includegraphics[width=80mm,height=50mm]{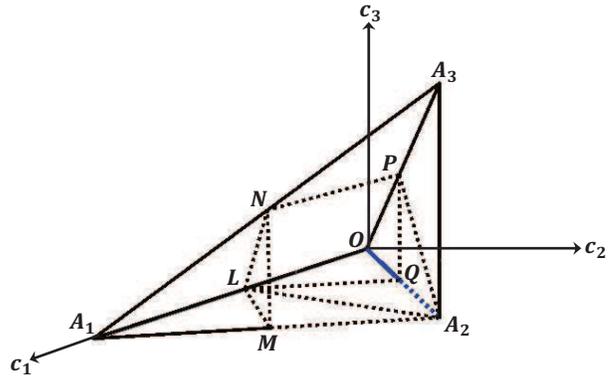}
\caption{(Color online) Tetrahedral ($OA_{1}A_{2}A_{3}$) representation of non-local two-qubit operations, known as the Weyl chamber. The points $L$, $M$, $N$, $P$ and $Q$, respectively, are the midpoints of the line segments $A_1O,  A_1A_2,  A_1A_3,  A_3O$, and $A_2O$ with $A_{1}=(\pi, 0, 0)$, $A_{2}=(\frac{\pi}{2}, \frac{\pi}{2}, 0)$, $A_{3}=(\frac{\pi}{2}, \frac{\pi}{2}, \frac{\pi}{2})$. Every point in the Weyl chamber corresponds to a local equivalence class of non-local two-qubit operations. The polyhedron $LMNPQA_2$ corresponds to perfect entanglers in the Weyl chamber. Line $LA_2$ identifies special perfect entanglers. The two-qubit gate $U(C_{0})$ belongs to the edge $A_2O$ illustrated in blue. Thus, the $U(C_{0})$ is a perfect entangler when it lies along the line segment $A_2Q$, which here corresponds to $\pi/8 \le\theta\le\pi/4$. In particular, for $\theta=\pi/4$ the gate $U(C_{0})$ is represented by the point $A_2$ in the Weyl chamber, which is the DCNOT local equivalence class of special perfect entanglers with local invariants $G1=0$, $G2=-1$ and therefore a maximum entangling power of $2/9$.}
\label{fig:weyl-chamber}
\end{figure}

The notion of local invariants introduced in Ref.~\cite{makhlin02} provides a novel geometric approach to uniquely characterize the local equivalence class of non-local two-qubit gates. For  $U(C_{0})$, we obtained the local invariants
\begin{eqnarray}
G_{1}&=&\frac{1}{4}[1+\cos(4\theta)]^2\nonumber\\
G_{2}&=&1+2\cos(4\theta).
\end{eqnarray}
From this pair of values, one can extract the symmetry reduced geometric coordinates $(c_{1}, c_{2}, c_{3})$ of the 3-Torus projected in a tetrahedron known as the Weyl chamber \cite{zhang03} (see Fig.  \ref{fig:weyl-chamber}). Each point in the Weyl chamber corresponds to a local equivalence class of non-local two-qubit operations. The three coordinates corresponding to the two-qubit operation $U(C_{0})$ read
\begin{eqnarray}
(c_{1}, c_{2}, c_{3})=(2\theta, 2\theta, 0),
\end{eqnarray}
which follow from the relations \cite{zhang03, rezakhani03}
\begin{eqnarray}
G_{1} &=& \frac{1}{4}[e^{-2ic_{3}}\cos2(c_{1} - c_{2}) + e^{2ic_{3}}\cos2(c_{1} + c_{2})]^{2}\nonumber\\
G_{2} &=& \cos(4c_{1}) +\cos(4c_{2}) +\cos(4c_{3}).
\end{eqnarray}
As shown in Fig. \ref{fig:weyl-chamber}, the two-qubit operation $U(C_{0})$ forms the edge $A_{2}O$ of the Weyl chamber for different values of $\theta$. 

\begin{figure}[h]
\centering
\includegraphics[width=70mm,height=45mm]{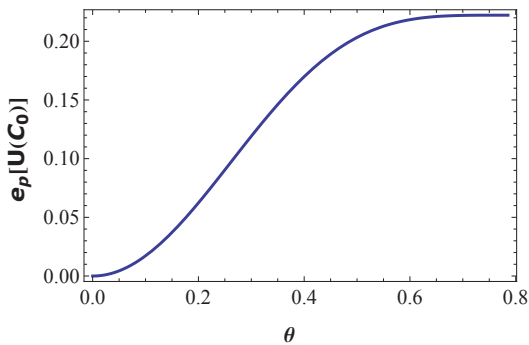}
\caption{(Color online)  Entangling-power $e_{p}[U(C_{0})]$ as a function of the control parameter $\theta$. For perfect entanglers $\frac{1}{6}\le e_{p}\le\frac{2}{9}$, which corresponds to $\pi/8\le\theta\le\pi/4$. Maximum entangling power of $2/9$ is achieved by special perfect entanglers at $\theta=\frac{\pi}{4}$.}
\label{fig:entangling-power}
\end{figure}

Moreover, the entangling capability of $U(C_{0})$ can be quantified by the entangling power \cite{zanardi00-r}, which is evaluated as \cite{balakrishnan09, balakrishnan10}
 \begin{eqnarray}
e_{p}[U(C_{0})]=\frac{2}{9}[1-|G_{1}|]=\frac{1}{18}[3-\cos^{2}(4\theta)-2\cos(4\theta)].\nonumber\\
\end{eqnarray}
In Fig. \ref{fig:entangling-power}, we depict the entangling-power of the $U(C_{0})$ gate as a function of the control parameter $\theta$. 

We conclude this section with two remarks on the entangling nature of the two-qubit gate $U(C_{0})$ in Eq. (\ref{eq:entangler}).
First, from the above analysis and illustrations in Figs. \ref{fig:weyl-chamber}, and \ref{fig:entangling-power}, we note that the two-qubit unitary operation $U(C_{0})$ is an entangling gate between the two register spin qubits for any values of $\theta$ except $\theta=0$. Furthermore, for $\pi/8\le\theta\le\pi/4$, the unitary $U(C_{0})$ belongs to the polyhedron within the Weyl chamber, which classifies the perfect entanglers, i.e., operators capable of producing maximally entangled state from some input product state \cite{zhang03}. In particular, when $\theta=\pi/4$, the $U(C_{0})$ gate has maximum entangling power of $2/9$ and corresponds to the vertex $A_{2}$ of the Weyl chamber, which represents a local equivalence class of special perfect entanglers, i.e., perfect entanglers that can maximally entangle a full product basis \cite{rezakhani03}. In fact the $U(C_{0}; \theta=\pi/4)$ is equivalent to the DCNOT gate, which is as efficient as the CNOT gate in quantum algorithms \cite{zhang04a, zhang04l}. 

The second remark concerns that the entangling characteristics of the geometric two-qubit gate $U(C_{0})$ only depend on the parameter $\theta$. This fact, together with Eq. (\ref{exchange-coupling-constants}), implies that any entangling power can be produced only with real exchange coupling constants $\alpha_{1}$ and $\alpha_{2}$. Thus, a two-qubit gate with arbitrary entangling power can be achieved only with the real part of the effective Hamiltonian in Eq. (\ref{eq:eff-H}), i.e., the anisotropic XY interaction Hamiltonian $H_{\text{XY}}$. 

\section{geometric interpretation}
\label{geometric interpretation}
Here, we discuss the geometric nature of the quantum gate in Eq. (\ref{eq:entangler}), which manipulates and entangles the register two-qubit system. As seen in the Sec.~\ref{Two-qubit entangling gate}, this gate is induced by evolving the four-dimensional conditional subspace $\mathcal{H}_{0}=\ket{0}\otimes\mathcal{H}_{r}$ around a path $C_{0}: [0, \tau]\ni t \rightarrow \mathcal{H}_{0}(t)$ governed by the time-dependent Schr\"odinger equation in such a way that $\mathcal{H}_{0}(\tau)=\mathcal{H}_{0}(0)=\mathcal{H}_{0}$. At each time, the subspace $\mathcal{H}_{0}(t)$ is a  four-dimensional subspace of  the eight-dimensional three qubit Hilbert space $\mathcal{H}$. Thus, the closed path $C_{0}$ resides in the Grassmann manifold $\mathcal{G}(8,4)$, i.e., the space of 4-dimensional subspaces of the 8-dimensional Hilbert space $\mathcal{H}$, which is in one-to-one correspondence with the set of all projections on $\mathcal{H}$ of rank $4$.

\begin{figure}[t]
\centering
\includegraphics[width=45mm,height=70mm]{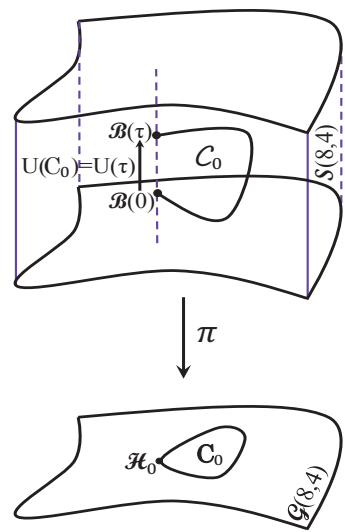}
\caption{(Color online) Schematic illustration of the principal bundle $\Gamma$ with total space $\mathcal{S}(8,4)$, base space $\mathcal{G}(8,4)$, and the fibers given by the unitary group $U(4)$. Each fiber represented by a blue vertical line in the total space is mapped via the natural projection $\pi$ to a point in the base space. For an initial choice of frame $\mathcal{B}(0)$ for $\mathcal{H}_{0}$,  the time-dependent Schr\"odinger equation lifts the closed path $C_{0}$ based at $\mathcal{H}_{0}$ in the Grassmann manifold $\mathcal{G}(8,4)$ into the unique horizontal curve $\mathcal{C}_{0}: [0, \tau]\ni t\rightarrow\mathcal{B}(t)$ in the Stiefel manifold $\mathcal{S}(8,4)$. In general, the horizontal lift is not closed and its two end points are connected by a holonomy element $U(\tau)$ of the principal bundle $\Gamma$ associated with the closed path $C_{0}$, with respect to the connection form $\mathcal{A}$.  The holonomy $U(\tau)$ is the two-qubit entangling gate $U(C_{0})$ when the inital frame $\mathcal{B}(0)$ is given by the register two-qubit computational frame.}
\label{fig:PB}
\end{figure}

Note that each state in $\mathcal{H}_{0}(t)$ along the cyclic evolution $C_{0}$ is linearly represented by a 4-frame, i.e., a set of four orthonormal basis state vectors of the corresponding subspace $\mathcal{H}_{0}(t)$. Hence, one may understand  the time evolution $C_{0}$ by studying how a given initial frame alters through this evolution. In other words, we need to examine the corresponding lifts of $C_{0}$ in the Stiefel manifold $\mathcal{S}(8,4)$, i.e.,  the space of all 4-frames in $\mathcal{H}$. As depicted in Fig. \ref{fig:PB}, the Stiefel manifold $\mathcal{S}(8,4)$ introduces a $U(4)$-principal bundle on $\mathcal{G}(8,4)$ denoted as $\Gamma=(\mathcal{S}(8,4), \mathcal{G}(8,4), \pi, U(4))$ with the natural projection
\begin{eqnarray}
\pi: \mathcal{S}(8,4)\rightarrow\mathcal{G}(8,4),
\end{eqnarray}
which maps each 4-frames to the corresponding 4-dimensional space spanned by that frame \cite{bohm2003, dariusz2004}.

Consider the lift
 \begin{eqnarray}
\mathcal{C}_{0}: [0, \tau]\ni t\rightarrow\mathcal{B}(t)
\end{eqnarray}
of $C_{0}$ in the Stiefel manifold $\mathcal{S}(8,4)$ specified by the time-dependent Schr\"odinger equation.
Here, at each time $\mathcal{B}(t)=\{\Psi_{a}(t), a=1, ..., 4\}$ is a 4-frame in $\mathcal{H}_{0}(t)$, where
\begin{eqnarray}
H_{\text{eff}}(t)\Psi_{a}(t)=i\hbar\frac{d\Psi_{a}(t)}{dt}
\label{SE}
\end{eqnarray}
for an initial choice of $\mathcal{B}(0)$. In order to examine the evolution of the initial frame $\mathcal{B}(0)$ about the lift $\mathcal{C}_{0}$, we assume $\tilde{\mathcal{B}}(t)=\{\tilde{\Psi}_{a}(t), a=1, ..., 4\}$ to be another once differentiable family of orthonormal
ordered basis of $\mathcal{H}_{0}(t)$ along $C_{0}$, such that $\tilde{\Psi}_{a}(\tau)=\tilde{\Psi}_{a}(0)=\Psi_{a}(0)$. Since both $\mathcal{B}(t)$ and $\tilde{\mathcal{B}}(t)$ are orthonormal bases of the same linear space $\mathcal{H}_{0}(t)$, there exists a unitary $U(t)\in U(4)$ such that 
\begin{eqnarray}
\mathcal{B}(t)=U(t)\tilde{\mathcal{B}}(t),
\end{eqnarray}
i.e.,
\begin{eqnarray}
\Psi_{a}(t)=\sum_{b=1}^{4}U_{ba}(t)\tilde{\Psi}_{b}(t).
\label{state-se}
\end{eqnarray}
for $a=1, ..., 4$. Therefore, at the end of the evolution we have   
\begin{eqnarray}
\mathcal{B}(\tau)=U(\tau)\tilde{\mathcal{B}}(\tau)=U(\tau)\tilde{\mathcal{B}}(0)=U(\tau)\mathcal{B}(0),
\label{eq:basis-se}
\end{eqnarray}
which indicates how a given initial frame $\mathcal{B}(0)$ evolves into a final frame $\mathcal{B}(\tau)$ about the lift $\mathcal{C}_{0}$ and consequently about the loop $C_{0}$. Explicitly, the Eq. (\ref{eq:basis-se}) shows that although the path $C_{0}$ is closed in $\mathcal{G}(8,4)$, the corresponding lift $\mathcal{C}_{0}$ is not necessarily closed in $\mathcal{S}(8,4)$ and its two end points in general are connected by a unitary element $U(\tau)$. By substituting Eq. (\ref{state-se}) in Eq. (\ref{SE}), we obtain 
\begin{eqnarray}
U(\tau)=\mathcal{T}\exp({i\int_{0}^{\tau}({\bf A}(t)-{\bf D}(t))dt}),
\label{C0-EV}
\end{eqnarray}
where 
\begin{eqnarray}
{\bf A}_{ab}(t)&=&i\bra{\tilde{\Psi}_{a}(t)}\frac{d}{dt}\ket{\tilde{\Psi}_{b}(t)},\nonumber\\
{\bf D}_{ab}(t)&=&\frac{1}{\hbar}\bra{\tilde{\Psi}_{a}(t)}H_{\text{eff}}(t)\ket{\tilde{\Psi}_{b}(t)}
\end{eqnarray}
and $\mathcal{T}$ denotes the time-ordering operator.

Eq. (\ref{C0-EV}) shows that the unitary operator $U(\tau)$ is in general composed of two parts: dynamical, which depends merely on the system Hamiltonian and is given by the phase factor ${\bf D}$, and geometrical, which is given by the gauge potential ${\bf A}$ \cite{anandan88, bohm2003}. The block off-diagonal form of the Hamiltonian  $H_{\text{eff}}(t)$ in Eq. (\ref{t-Hamiltonian}) implies that 
\begin{eqnarray}
P_{0}(t)H_{\text{eff}}(t)P_{0}(t)=0,
\end{eqnarray}
where $P_{0}(t)$ is the projection operator on $\mathcal{H}_{0}(t)$, and in fact ${\bf D}=0$. This means, no dynamical phases occure in the cyclic evolution $C_{0}$. Therefore, the transformation $U(\tau)$ is fully determined by the geometric part and can be written as a path-ordered integral 
\begin{eqnarray}
U(\tau)=\mathcal{P}\exp({i\oint_{C_{0}}\mathcal{A}}),
\end{eqnarray}
where $\mathcal{A}(t)={\bf A}(t)dt$ introduces a connection one-form on the Grassmann manifold $\mathcal{G}(8,4)$ \cite{bohm2003}.  This in turn implies that indeed the lift $\mathcal{C}_{0}$ identifies the unique horizontal lift of $C_{0}$ started at $\mathcal{B}(0)$ and the unitary operation $U(\tau)$ is the holonomy of $\mathcal{C}_{0}$ with respect to the connection form $\mathcal{A}$ \cite{anandan88, bohm2003, dariusz2004}. In fact, for any reference point $\mathcal{B}(0)\in\pi^{-1}(\mathcal{H}_{0})\subseteq\mathcal{S}(8,4)$, the $U(\tau)$ is a holonomy element of the principal bundle $\Gamma$ associated with the closed path $C_{0}$. Especially, if $\mathcal{B}(0)$ is the register two-qubit computational basis,
\begin{eqnarray}
U(\tau)=U(C_{0}),
\end{eqnarray}
which confirms that the two-qubit entangling gate $U(C_{0})$ is a holonomy element of the principal bundle $\Gamma$. Therefore, $U(C_{0})$ is geometric in nature and fully determined by the geometric structure of the principal bundle $\Gamma$ and the loop $C_{0}$. (see Fig. \ref{fig:PB}).

We end this section with a discussion on the practical validity of the geometric entangling gate $U(C_{0})$. The crucial point in achieving the gate $U(C_{0})$ is to evolve the subspace $\mathcal{H}_{0}$ in a cyclic fashion. This relies on the ability of manipulating the exchange parameters $J_{k}$ and $D_{k}^{z}$, $k=1,2$ with the same time-dependent function $\Omega(t)$. From the Hubbard model description of a spin chain, the exchange parameters $J_{k}$ and $D_{k}^{z}$, respectively, associate with the spin-independent and spin-dependent hopping between the corresponding sites \cite{trif10}. They are distinguished and considered separately because one is spin-dependent and usually smaller than the spin-independent one, but the microscopic origin of them is tunneling through a potential barrier.  Thus,  it is quite reasonable that, in the first approximation, when this potential barrier is varied in time by applying an external gate voltage, the time dependence of these parameters
should be the same. This implies that the overall time-dependence of $J_{k}$ and $D_{k}^{z}$ is indeed the same.
However, as expressed in Sec. \ref{Two-qubit entangling gate}, the two-qubit gate $U(C_{0})$ with arbitrary entangling power can be achieved only with the interaction Hamiltonian $H_{\text{XY}}$. Explicitly, one can only use $H_{\text{XY}}$ as the effective Hamiltonian to perform the two-qubit entangling gate $U(C_{0})$ and consider the Dzyalozhinsky-Moriya term as a decoherence effect in the system. In this way, only the exchange parameters $J_{k}$s need to be controlled. Moreover, since $H_{\text{XY}}$ has exactly the same block off-diagonal form as the Hamiltonian $H_{\text{eff}}(t)$ in Eq. (\ref{t-Hamiltonian}), the gate carried out with $H_{\text{XY}}$ would as well be geometric in nature. 

\section{Robustness}
\label{Robustness}
Geometric phases and quantum holonomies depend only on global properties of quantum evolutions and geometric structure of the state space \cite{bohm2003, dariusz2004}.  Thus, they are inherently resilient to local perturbations, external parameter noises, and some other class of errors associated with specific details of how the evolutions are carried out \cite{Pachos2001, sjoqvist2008, Solinas2012}. Namely, quantum gates based upon quantum holonomies have some built-in fault-tolerant features and stability, which can be employed to achieve robust quantum computation. Among geometric gates, nonadiabatic non-Abelian geometric gates may have some additional advantages, such as they are exact in a sense that there is no adiabatic approximation, there is no need for slow manipulation and in fact there is more freedom in the gates operation times \cite{sjoqvist16}. Therefore, one may expect that nonadiabatic geometric gates can be made robust to wider class of noises compare to their adiabatic counterpart. The robustness of nonadiabatic holonomic gates against some general sources of errors has been studied in Ref. \cite{Johansson12}. Below, we evaluate the fidelity of the geometric entangling gates demonstrated in the former sections. In this evaluation, we examine the special perfect entangler performed with $\theta=\frac{\pi}{4}$, i.e., $U(C_{0}; \theta=\frac{\pi}{4})$, as a test gate against some source of perturbations and decoherences.

\begin{figure}[h]
\centering
\includegraphics[width=85mm,height=60mm]{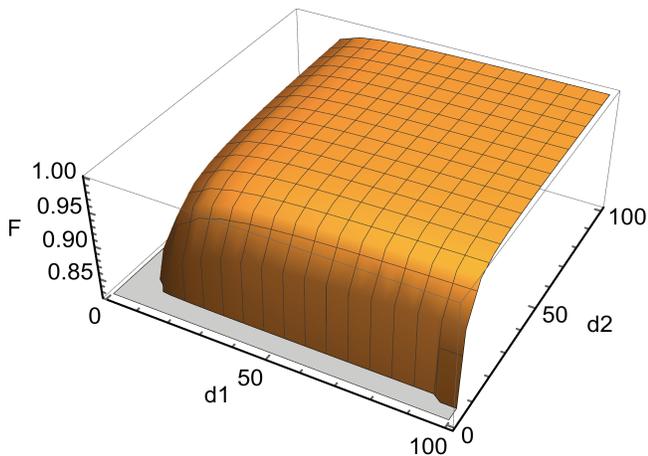}
\caption{(Color online) Fidelity, $F$, of the special perfect entangler $U(C_{0}; \theta=\frac{\pi}{4})$, which is carried out only with the XY interaction Hamiltonian term $H_{\text{XY}}$ in Eq. (\ref{eq:eff-H}), against dimensionless parameters $d_{i}=\frac{2\Omega\omega}{\hbar D^{z}_{i}}=\frac{\sqrt{|J_{1}|^{2}+|J_{2}|^{2}}}{D^{z}_{i}}$, $i=1, 2$, corresponding to the Dzyalozhinsky-Moriya spin-orbit interaction. Here, we have used a square pulse function with amplitude $\Omega$ for the scaling function $\Omega(t)$.}
\label{fig:fidelity-DM}
\end{figure}

As discussed above, the $H_{\text{XY}}$ can be used as the effective Hamiltonian to perform the two-qubit entangling gate $U(C_{0})$ and consider the Dzyalozhinsky-Moriya term as a perturbative noise in the system. In this way, only the exchange parameters $J_{k}$s need to be controlled. 
Figure \ref{fig:fidelity-DM} illustrates the fidelity of the special perfect entangler $U(C_{0}; \theta=\frac{\pi}{4})$, which is implemented through the XY interaction Hamiltonian $H_{\text{XY}}$, against Dzyalozhinsky-Moriya spin-orbit interaction contribution to the system. It is important to mention that a dominating source of noise in spin systems is the spin-orbit interaction \cite{glazov2004, russ2016, prada2017}. 

\begin{figure}[h]
\centering
\includegraphics[width=85mm,height=60mm]{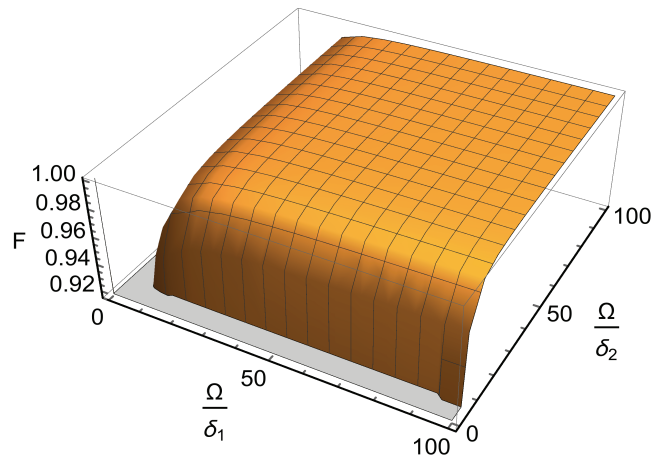}
\caption{(Color online) Fidelity, $F$, of the special perfect entangler $U(C_{0}; \theta=\frac{\pi}{4})$ against parameter noise. A square pulse function with amplitude $\Omega$ has been considered for the scaling function $\Omega(t)$. For parameter noises, we assume the pulse intensity $\Omega$ is perturbed independently on the two interacting arms in Fig. \ref{fig:3-spin-chain} coupling the auxiliary spin qubit to the two register qubits as $\Omega_{1}=\Omega+\delta_{1}$ and $\Omega_{2}=\Omega+\delta_{2}$. The fidelity is plotted as a function of dimensionless parameters $\frac{\Omega}{\delta_{i}}$, $i=1, 2$.}
\label{fig:fidelity-PN}
\end{figure}

Controlling the exchange parameters is of crucial importance to gate operation. To study the effect of parameter noise in gate operation, we assume a square pulse envelope with amplitude $\Omega$ for the scaling function $\Omega(t)$. We further assume that the pulse intensity $\Omega$ is perturbed independently on the two interacting arms in Fig. \ref{fig:3-spin-chain} coupling the auxiliary spin qubit to the two register qubits as $\Omega_{1}=\Omega+\delta_{1}$ and $\Omega_{2}=\Omega+\delta_{2}$, due to for instance imprecise control of the system parameters. The fidelity of the special perfect entangler $U(C_{0}; \theta=\frac{\pi}{4})$ versus this type of error is shown in Fig. \ref{fig:fidelity-PN}.

\begin{figure}[h]
\centering
\includegraphics[width=70mm,height=45mm]{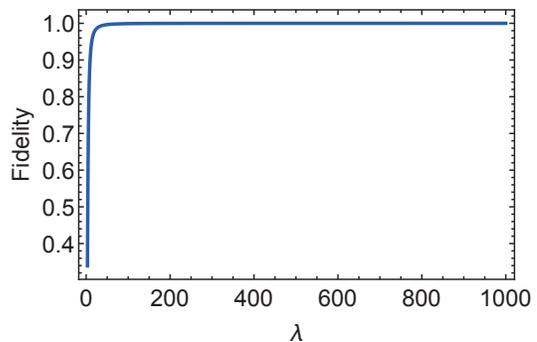}
\caption{(Color online) Fidelity of the special perfect entangler $U(C_{0}; \theta=\frac{\pi}{4})$ against dephasing due to the hyperfine interaction with nuclei. The fidelity is plotted as a function of dimensionless parameter $\lambda=\frac{\tau_{\text{hd}}}{\tau_{\text{op}}}=\frac{N}{\hbar A\tau_{\text{op}}}$, i.e., the ratio between the hyperfine decoherence time, $\tau_{\text{hf}}$, and the gate operation time, $\tau_{\text{op}}$. Here, we have considered a square pulse function with amplitude $\Omega$ for the scaling function $\Omega(t)$ and that each electron spin interacts with two nuclear spins homogeneously, when all the coupling constants are the same.} 
\label{fig:fidelity-dephasing}
\end{figure}

An important source of decoherence in a quantum dot electron spin qubit system is the problem of electron spin dephasing due to the hyperfine interaction with nuclei. We consider the three localized electron spins in the above setup interacting with nuclear spins via the Fermi contact hyperfine interaction. This interaction is described by the interaction term \cite{khaetskii2002, coish2004} 
\begin{eqnarray}
H_{\text{hi}}&=&{\bf h}^{(a)}\cdot {\bf S}^{(a)}+{\bf h}^{(1)}\cdot {\bf S}^{(1)}+{\bf h}^{(2)}\cdot {\bf S}^{(2)}
\end{eqnarray}
introduced into the Hamiltonian in Eq. (\ref{t-Hamiltonian}), where ${\bf h}^{(l)}=(h_{x}^{(l)}, h_{y}^{(l)}, h_{z}^{(l)})=\sum_{k=1}^{N}A_{l;k}{\bf I}^{(k)}$ is the quantum field generated by an environment of $N$ nuclear spins at electron lattice site $l=a, 1, 2$. Here, ${\bf S}^{l} = (S_{x}^{(l)}, S_{y}^{(l)}, S_{z}^{(l)})$ is the electron spin operator, ${\bf I}^{(k)} = (I_{x}^{(k)}, I_{y}^{(k)}, I_{z}^{(k)})$ is the nuclear spin operator at nuclear lattice site $k$ and $A_{l;k}$ is the associated hyperfine coupling constant.

Figure \ref{fig:fidelity-dephasing} shows the effect of dephasing on the special perfect entangler $U(C_{0}; \theta=\frac{\pi}{4})$ in the case of homogeneous
coupling, when all the coupling constants are equal, i.e. $A_{l;k} = A/N$. We have used nuclear spin $1/2$ and the operator-sum representation approach to plot the fidelity as a function of dimensionless parameter $\lambda=\frac{\tau_{\text{hd}}}{\tau_{\text{op}}}=\frac{N}{\hbar A\tau_{\text{op}}}$, which is the ratio between the hyperfine decoherence time, $\tau_{\text{hf}}=\frac{N}{\hbar A}$, and the gate operation time, $\tau_{\text{op}}$. Considering the $\tau_{\text{hf}}$ in the order of microsecond ($\mu s$) \cite{khaetskii2002, coish2004} would permit $99\%$ gate fidelity with $\tau_{\text{op}}<100\ ns$.

From the above analysis, we can conclude that a careful control of inter-dot tunnel couplings indeed allows high fidelity performance of the geometric entangling $U(C_{0})$-gates.

\section{summary}
\label{summary}
In summary, we have introduced a feasible setup to create geometric entanglement between spin qubits. We have used a system of three-body spin-chain, whose dynamic is described by the anisotropic XY interaction Hamiltonian plus an antisymmetric Dzyalozhinsky-Moriya spin-orbit interaction. In this system, two register spin qubits are coupled through an auxiliary spin qubit. We show that by electrical manipulation of inter-qubit exchange couplings, nonadiabatic holonomic two-qubit entangling gates between register qubits can be realized, provided the auxiliary qubit is initialized in the $\ket{0}$ state. Both the geometric and entangling natures of the proposed gates have been analyzed in detail. Our analyses show that the system allows for implementation of geometric two-qubit gates with any arbitrary entangling power. Moreover, it shows that by a careful control of exchange couplings, special perfect entanglers equivalent to DCNOT gate, which is as efficient as CNOT gate in quantum algorithm, can be achieved. It has also been shown that the entangling nature of the proposed gates depends only of the anisotropic XY interaction term. This indeed indicates that any geometric entangling power can be realized only with the anisotropic XY interaction Hamiltonian. We illustrate the fidelity of the special perfect entangler performed by only the anisotropic XY interaction Hamiltonian against Dzyalozhinsky-Moriya spin-orbit interaction contribution to the system. Furthermore, we examine the effect of the other types of decoherences like dephasing and parameter noises on this special perfect entangler. It turns out that a careful control of inter-qubit exchange couplings gives rise to high fidelity performance of the gate. The electrical, nonadiabatic, and geometric natures of the proposed entangling gates together provide a proper and feasible way to generate fast and robust entanglement between spin qubits.

\section*{Acknowledgments}
The author acknowledges support from the Department of Mathematics at
University of Isfahan, Iran.

\end{document}